\documentclass[runningheads]{llncs}
\usepackage[T1]{fontenc}
\usepackage{graphicx}
\usepackage{booktabs}
\usepackage[misc]{ifsym}
\newcommand{\corr}{(\Letter)}
\usepackage{mwe}
\usepackage{amssymb}
\usepackage{amsmath}
\usepackage[dvipsnames]{xcolor}
\newcommand{\correction}[1]{{#1}}
\usepackage{multirow}
\usepackage{appendix}
\usepackage{hyperref}
\usepackage{algorithm2e}

\begin{document}

\title{Learning local and global prototypes with optimal transport for unsupervised anomaly detection and localization}

\titlerunning{Prototype-based anomaly detection with optimal transport}

\author{Robin Trombetta\inst{} \corr \and
Carole Lartizien\inst{}
}

\authorrunning{R. Trombetta et al.}

\institute{Univ. Lyon, CNRS UMR 5220, Inserm U1294, INSA Lyon, UCBL, CREATIS, France \email{robin.trombetta@creatis.insa-lyon.fr}
}

\maketitle              

\begin{abstract}
Unsupervised anomaly detection aims to detect defective parts of a sample by having access, during training, to a set of normal, \textit{i.e.} defect-free, data. It has many applications in fields, such as industrial inspection or medical imaging, where acquiring labels is costly or when we want to avoid introducing biases in the type of anomalies that can be spotted.
In this work, we propose a novel UAD method based on prototype learning and introduce a metric to compare a structured set of embeddings that balances a feature-based cost and a spatial-based cost. We leverage this metric to learn local and global prototypes with optimal transport from latent representations extracted with a pre-trained image encoder.
We demonstrate that our approach can enforce a structural constraint when learning the prototypes, allowing to capture the underlying organization of the normal samples, thus improving the detection of incoherencies in images. Our model achieves performance that is on par with strong baselines on two reference benchmarks for anomaly detection on industrial images. \correction{The code is available at  \href{https://github.com/robintrmbtt/pradot}{https://github.com/robintrmbtt/pradot}.}

\end{abstract}

\section{Introduction}

Anomaly detection refers to the task of identifying samples that deviate from the distribution of normal cases. 
In domains such as video analysis or medical imaging, obtaining large-scale annotated datasets is often prohibitively expensive. Furthermore, relying on labeled data inherently biases the model toward detecting only known types of anomalies.
As a result, rare or previously unseen anomalies can go undetected. To address these limitations, unsupervised anomaly detection (UAD) has emerged as an alternative approach to supervised learning.
Instead of learning from both normal and abnormal examples, UAD models are trained exclusively on normal (\textit{i.e.}, anomaly-free) data and learn to capture the underlying distribution or structure of the normal data.
At test time, the model computes the distance between the new sample and the learnt data distribution. Based on this measure, it is expected to differentiate between in-distribution samples and samples that deviate from the anomaly-free distribution.
The research in the field on UAD for automatic industrial inspection was greatly pushed by the MVTec AD dataset \cite{bergmann2019_mvtecad}, which set a robust benchmark for the evaluation of UAD methods. More recently, a new version of the dataset called MVTec AD LOCO \cite{Bergmann2022_mvtecadloco} was released to address the shortcomings of the first benchmark. In particular, it introduces anomalies of a more logical nature, which violate the expected structural or semantic organization of the image, the detection of which can require capturing long-range dependencies between the different subparts of the image. 

To address the challenge of unsupervised anomaly detection, a wide range of methods have been developed, employing different strategies to localize abnormal regions in images.
A prominent family of methods relies on the extraction of hierarchical features from a pre-trained encoder to build a memory bank, that is, a set of discrete prototypical features representing normal patterns \cite{brendel2018_padim,cohen2020_spade,roth2022_patchore,guo2023_templateguided,park2020_mnad}. At inference time, a distance metric quantifies the discrepancy between the latent features of the test image and the learnt prototypes. The abnormal parts of the image are identified when there is a high dissimilarity between a latent feature and the prototypes. Various techniques allow to learn such prototypes such as kNN \cite{cohen2020_spade,guo2023_templateguided} or coreset sampling \cite{roth2022_patchore}. 
These methods have received particular attention because of their high performance, simplicity, speed and efficiency at low data regime. One of their shortcomings is their limited ability to spot abstract incoherencies in the images, such as misplaced objects, even when employing multi-level feature maps and improved neighbour-awareness mechanisms \cite{cohen2020_spade,roth2022_patchore}, 
\\

In this work, we propose using optimal transport to efficiently learn prototypes for features extracted from the pre-trained encoder. Inspired by fused Gromov-Wasserstein \cite{titouan2019_fgw} (FGW) distance, we introduce a discrepancy function that balances a feature-based cost and a spatial-based cost. By playing on the ratio between these two components, we are able to learn more \textit{local} or more \textit{global} representatives of the normal class, thus capturing diverse patterns on the training images. By applying it to features extracted from a pre-trained backbone, such as ResNet50 \cite{he2016resnet}, we propose a novel method for Prototype-based Anomaly Detection with Optimal Transport (PRADOT). Compared to previous state-of-the-art methods that leverage a memory bank \cite{cohen2020_spade,roth2022_patchore}, our approach, in particular due to the structural constraint enforced by assignment to local prototypes, is shown to better identify incoherencies in images. We evaluate our method on MVTec AD and MVTec AD LOCO, two standard benchmarks for unsupervised anomaly detection and localization on industrial images, where it achieves competitive performances compared to state-of-the-art methods. 

\section{Related Work}

\subsubsection{Unsupervised anomaly detection.} Earliest works focusing on anomaly detection tried to estimate the support of the normal samples with one-class support machine or support vector data description \cite{tax2004svdd,ruff18a_deepsvdd}. A large family of methods makes use of generative models, in particular Auto-Encoders or Generative Adversarial Nets \cite{goodfellow2020gan,schlegl2019fanogan}, and expects to localize anomalies with the pixel-wise distance between an input image and its reconstruction. 
However, recent studies have highlighted the fact that auto-encoders can generalize their reconstruction abilities to anomalous parts of images \cite{zavrtanik2021reconstructionlimit}, 
thus limiting the effectiveness of such approaches. To overcome this issue, more advanced auto-encoder based methods have incorporated masking strategies \cite{zavrtanik2021reconstructionlimit,yan2021maskgan}, memory mechanisms \cite{park2020_mnad,gong2019memoryae} or synthetic data \cite{zavrtanik2021draem,schluter2022naturalsynthetic}. Recently, the generative abilities of diffusion models \cite{ho2020ddpm} have also been employed for unsupervised anomaly detection \cite{yao2024glad}.

Many UAD methods exploit the representative power of features extracted from pre-trained encoders. 
Density-based methods learn the distribution of normal features via direct likelihood estimation \cite{brendel2018_padim,li2021cutpaste} while support-based methods \cite{ruff18a_deepsvdd,yi2020patchsvdd} aim to spot anomalies as samples with a high distance to the estimated boundaries of the distribution of anomaly-free cases. Normalizing flows have been widely applied in UAD, mainly through network-parametrized density estimation of deep latent features \cite{yu2021fastflow,zhou2024_msflow}. In knowledge distillation, a frozen pre-trained encoder is used as a teacher to guide the training of a student network, and the anomalies are detected by a large discrepancy between the features of the two networks \cite{wang2021studentteacherpyramid,Deng2022_reverseKD,zhang2024contextualknowledgedistillation}. 
\correction{The methods most closely related to this work are those based on prototypes \cite{cohen2020_spade,park2020_mnad,roth2022_patchore,guo2023_templateguided}. A cornerstone work falling in this category is PatchCore \cite{roth2022_patchore}, which builds a memory bank of locally aware prototypes with coreset sampling and uses nearest neighbours search at inference to evaluate the distance between a feature and the discrete representatives of the normal samples.}

\subsubsection{Prototype learning.} Prototypical-based methods are popular in machine learning, one of their advantages being that they provide good explainability. Indeed, their exemplar-driven decision process generally allows for intuitive interpretation. One of the most popular applications of prototype learning is the Vector-Quantized Variational Auto-Encoder (VQ-VAE) \cite{van2017vqvae}, where the bottleneck of a typical Auto-Encoder is replaced by a dictionary for discrete encoding. SwAV \cite{caron2020swav} is a widespread self-supervised technique that makes use of assignments to discrete clusters to learn visual features. \cite{zhou2022rethinking_proto} analyses the task of semantic segmentation from a prototype view and explicitly learns multiple prototypes of pixel embeddings to better capture intra-class diversity. This paradigm also found applications in few- and zero-shot learning \cite{snell2017prototypicalnetworks_fewshot,kim2019variationalprotoenc}, sample-level out-of-distribution detection \cite{lu2024mixtureot_ood} or unsupervised anomaly detection \cite{roth2022_patchore,guo2023_templateguided,park2020_mnad}. 

\subsubsection{Optimal transport.} 
Optimal transport (OT) provides a mathematical framework for comparing probability distributions by identifying the minimal cost required to shift from one to the other.
Computational methods for OT have successfully been applied to many applications in machine learning such as domain adaptation \cite{courty2016optimal_domain_adapt,damodaran2018deepjdot}, visual place recognition \cite{izquierdo2024visualplacerecognition}, zero-shot segmentation \cite{kim2023zegot}, multi-modal problems \cite{xu2023multimodal_ot_mil,cao2022otkge} or out-of-distribution detection \cite{lu2024mixtureot_ood,lu2023characterizing_ood_ot}. In methods that use prototypes or clusters, optimal transport is often used for assignment during training \cite{caron2020swav,zhou2022rethinking_proto}. It can force equipartition between the discrete representatives and thus avoid having unassigned prototypes or collapsing prototypes. \correction{Similar to our work, optimal transport is used in \cite{shan_fewshotadot,tian2025fastref_fewshot} for industrial anomaly detection, in their case in a few-shot learning setup. \cite{shan_fewshotadot} generates bilateral semantic correspondence between a query image and a small guidance set, while \cite{tian2025fastref_fewshot} refines a set a prototypes by computing the optimal transfer plan with a query image.} When employed on large-scale data, the classical OT problem is often reformulated by adding an entropic regularization term, which allows much faster computation \cite{cuturi2013_entropy}. Through (fused) Gromov-Wasserstein \cite{titouan2019_fgw,peyre2016gromov}, optimal transport is often used in graph-related problems \cite{xu2024ot_unsupervisedactionseg,thual2022aligning_ufgw,vincent2022template_graph_ot}.

\section{Method}

\begin{figure}
     \centering
     \includegraphics[width=\linewidth]{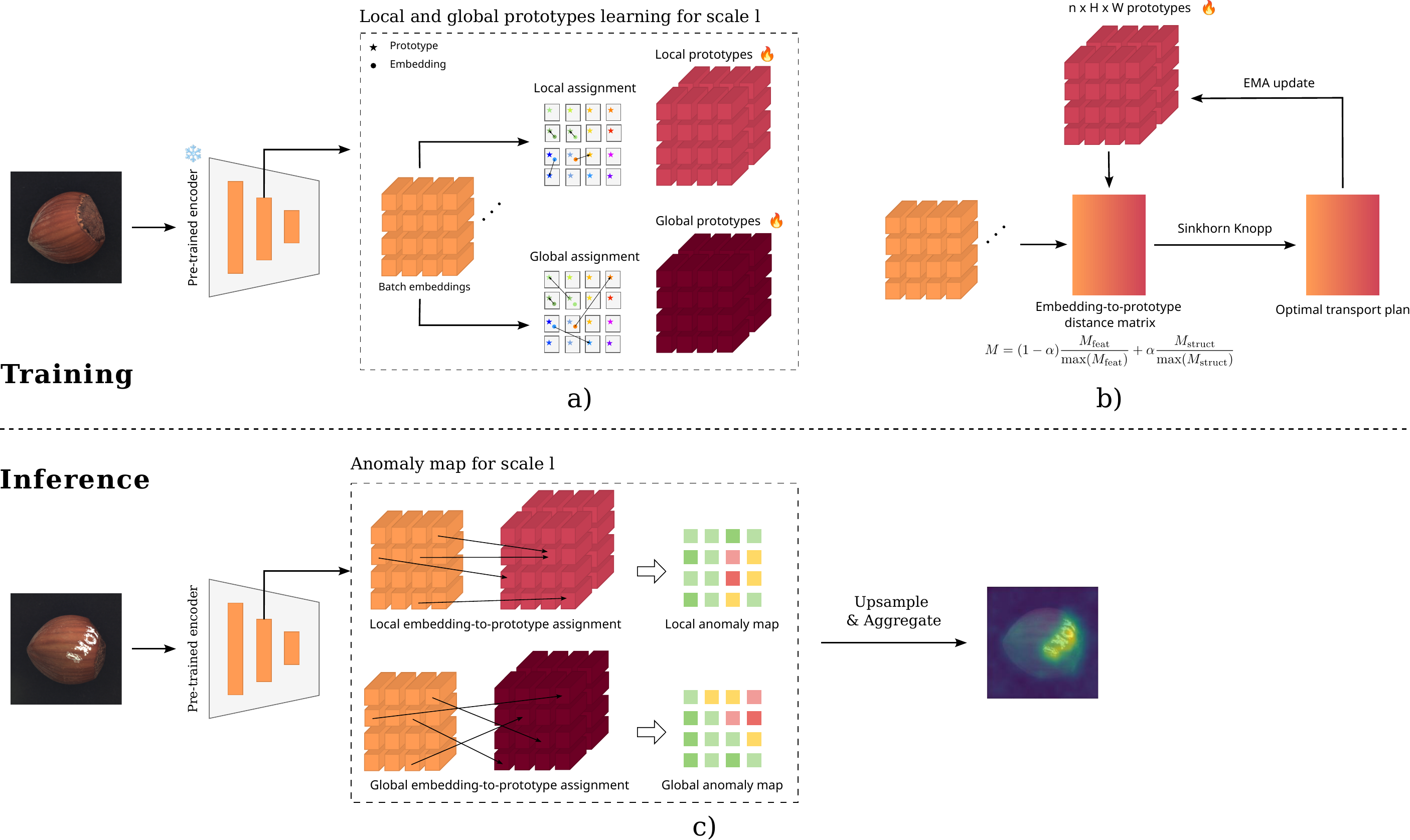}
     \caption{Overview of the proposed method. a) During training, we extract features from the training images and learn local and global prototypes with b) optimal transport assignment from embeddings to prototypes and exponential moving average update. c) At inference, the distances between each of the image embeddings and the closest local and global prototypes serve as anomaly score.}
     \label{fig:method_overview}
 \end{figure}

Given a set $\mathcal{D}^{\text{train}} = \{  (x_i,y_i) \}_{i=1}^{N_{\text{train}}}$ of normal images $x_i$ ($y_i = 0)$,
and a test set $\mathcal{D}^{\text{test}} = \{ 
 (x_j, y_j) \}_{j=1}^{N_{\text{test}}}$ containing normal and abnormal images $x_j$ with their associated pixel-level ground-truth $y_j$, we aim to train a model solely on normal images which is able to detect whether or not a test image lies inside the normal distribution and if not localize where are the abnormal parts of the image. The overview of the proposed method is illustrated in Figure \ref{fig:method_overview}, and the method is described in detail in the following sections.
 
 Its principle can be summarized in three main steps: 1) We extract multi-scale features from the training images with a pre-trained encoder 2) Using optimal transport with a tailored cost function, we assign each feature of a batch to the local and global prototypes and update the prototypes with exponential moving average 3) At inference, we assign each feature vector extracted from the input image to the closest prototype and use the distance between the two as indicator of the presence or absence of an anomaly.

\subsection{Prototype learning with optimal transport}

 \subsubsection{Hierarchical feature extraction.}

 As many anomaly detection methods, we consider a pre-trained encoder $\phi_\theta$, typically a ResNet50, that extracts multi-level features $\phi_\theta (x) = \{ z^l \}_l$ from an input image $x$, where $l$ is the stage of the encoder at which the features are extracted and $z^l \in \mathbb{R}^{H_l \times W_l \times D_l}$ is the deep representation of the input and composed of $H_l \times W_l$ feature vectors of dimension $D_l$. 
Given a set of intermediate features $\mathcal{F}^l = \{ z_i ^l \}_{i=1}^{N_{\text{train}}}$ extracted from the images of the training set, the idea is to leverage optimal transport to learn the barycentre of the features $\mathcal{F}_l$. In many applications, images have a local structure and coherence, hence we want to design a way to capture local information and structure from the training dataset. For simplicity and as the method is independent of the stage $l$ of the encoder where the features are extracted, we will drop this variable from the notations in the following. 

\subsubsection{Problem setup.}
We can consider that the features $z$ extracted by the encoder $\phi_\theta$ from an input image $x$ can be divided in a regular grid of size $H \times W$ and consequently embed each feature vector at the position $(i,j)$ of this grid with its (normalized) coordinates $c_{ij} = (\frac{i}{H}, \frac{j}{W})$, with $(i,j) \in [\![ 1,H ]\!] \times [\![ 1,W ]\!]$. From now on, the term \textit{input embedding} will refer to both the feature vector extracted from the encoder and its coordinates, \textit{i.e.} $f_{ij} = (z_{ij}, c_{ij}) \in \mathbb{R}^{D} \times \mathbb{R}^2$.

We define a set of prototypes $\mathcal{P} = \{ P_i \}_{i=1} ^{N_{p}}$ where a prototype $P_{i} = (p_i, \rho_i)$ is composed of a feature vector $p_i \in \mathbb{R}^{D}$ and spatial coordinates $\rho_i = (\frac{\rho_{x,i}}{H}, \frac{\rho_{y,i}}{W}) \in \mathbb{R}^2$. Although in theory the coordinates of the prototypes could be left free, we decide to simplify the modelling by imposing that the prototypes be placed on the same grid as the embeddings. Hence, their coordinate vectors $(\rho_{x,i},\rho_{y,i})$ are defined only on $[\![ 1,H ]\!] \times [\![ 1,W ]\!]$. To allow for a richer representation, we set that there can be $n$ prototypes per grid element, resulting in a total number of prototypes of $ N_p = n \times H \times W$.


To quantify the closeness between an input embedding $f=(z,c)$ and a prototype $P=(p,\rho)$, we define the following cost, inspired by the fused Gromov-Wasserstein distance \cite{titouan2019_fgw}:
\begin{equation}
\label{eq:cost}
    \mathcal{C}(f,P) = (1 - \alpha)(1 - \frac{\langle z, p\rangle}{\| z \| \| p \|}) + \alpha \| c - \rho\|^2
\end{equation}
where $\alpha \in [0,1]$ is a hyperparameter and $\langle \cdot, \cdot \rangle$ designates the dot product between vectors. This cost introduces a balance between a feature-based cost, with the cosine distance between the feature vectors of the image embedding and the prototype, and a structural (or spatial) cost with the $L^2$ norm between their 2D coordinates.

\subsubsection{Optimal transport problem.}
Given a set of input images $\{x_i\}_{i=1}^N$, we formalize an optimal transport problem between their extracted embeddings $\{f_{ij}^n\}_{i,j,n=1}^{i=H,j=W,n=N}$ and the prototypes $\{ P_i\}_{i=1}^{N_p}$ as the search for a transport plan $T^*$ which minimizes the total transportation cost between embeddings and prototypes, while respecting the total mass constraint:
\begin{equation}
\label{eq:OT}
    \begin{aligned}
    T^* = \min_{T \in \Pi(\mu, \nu)} \langle M, T \rangle = \min_{T \in \Pi(\mu, \nu)} \sum_{i=1}^{N\times H \times W} \sum_{j=1}^{N_p} M_{ij} T_{ij} \\
    \textrm{s.t } \Pi(\mu, \nu) = \{ T \in \mathbb{R}_{+}^{(NHW)\times N_p} | T \textbf{1}_{N_p} = \mu, ^{t}T \textbf{1}_{NHW} = \nu \},
    \end{aligned}
\end{equation}
where $\langle \cdot, \cdot \rangle$ designates the Frobenius dot product. $\mu$ and $\nu$ are the mass distributions respectively associated with the embeddings and the prototypes; without any \textit{a priori} knowledge on the data, both are set to uniform distributions. $M \in \mathbb{R}^{(NHW) \times N_p}$ designates the total cost matrix between the input embeddings and the prototypes. More specifically, this cost matrix is a normalized version of the cost $\mathcal{C}$ introduced in Eq. (\ref{eq:cost}):
\begin{equation}
\label{eq:cost_M}
    M = (1 - \alpha) \frac{M_{\text{feat}}}{\max (M_{\text{feat}})} + \alpha \frac{M_{\text{struct}}}{\max (M_{\text{struct}})}
\end{equation} 
with $M_{\text{feat}}(i,j) = 1 - \frac{\langle z_i, p_j\rangle}{\| z_i \| \| p_j \|}$ and $M_{\text{struct}}(i,j) = \| c_i - \rho_j \|^2$. \\

In practice, solving Eq. (\ref{eq:OT}) is not computationally efficient 
In our case, considering ResNet50 as backbone and input images of size $224 \times 224$, for $l=2$, a grid of features is composed of $28 \times 28= 786$ vectors. If we consider solving the problem for multiple images and prototypes, the size of the cost matrix will rapidly grow and the solution will be intractable. To overcome this issue, we consider instead the entropic version of the optimal transport problem \cite{cuturi2013_entropy} by introducing a regularization term $H(T)$:
\begin{equation}
\label{eq:ot_entropy}
    \min_{T \in \Pi(\mu, \nu)} \langle M, T \rangle - \epsilon H(T),
\end{equation}
where $\epsilon > 0$ and $H(T) = \sum_{i,j} T_{ij}(\log T_{ij} - 1)$. This formulation makes the problem strictly convex, which allows for much faster solving using the Sinkhorn-Knopp algorithm. \\

\subsubsection{Prototypes update.} 
Instead of computing the optimal transport plan between the prototypes and the entire training dataset at once, we process the inputs in batches. 
This strategy not only enables the method to scale to large datasets but also allows for the use of data augmentation during training, thereby increasing sample diversity and reducing the risk of overfitting.
At each step of the training, we draw a subset of images $\{ x_i\}_{i=1}^B$ from our training set, with $B \geq n$, and find the optimal transport $T^*$ plan in Eq. (\ref{eq:ot_entropy}) between all $B$ embeddings of the considered batch $\{f_{ij}^n\}_{i,j,n=1}^{i=H,j=W,n=B}$ and all $n \times H \times W$ prototypes. $T^*$ contains the assignment weights between batch embeddings and prototypes. Analogous to the barycentre search algorithm in \cite{titouan2019_fgw}, the features of the prototypes are updated using these assignation weights:

\begin{equation}
    p_i \xleftarrow{} \eta p_i + (1 - \eta) N_p \sum_{k=1}^{B\times H \times W} T^*(k,i)z_k
\end{equation}
where $\eta \in [0,1]$ is the exponential moving average update parameter. When the parameter in the cost $\alpha$ equals zero, the prototypes are only feature-based (or \textit{global}) barycentres of the training dataset. As $\alpha$ increases, a prototype $p_i$ gathers information from its spatial neighbours and becomes more of a \textit{local} barycentre. \correction{The pseudocode describing prototype learning is provided in Appendix \ref{app:pseudo_code}.}

\subsection{Anomaly localization et image-level score} Once the prototype features have been learnt, we can infer the anomaly map for defect localization as well as the image-level score. First, we pass the image through the encoder $\phi_\theta$ to get their intermediate feature representation $\{ F^l\}_l$. For each embedding $f=(z,c)$, the anomaly score $S_f$ is the minimum distance between the embedding and the prototypes for the cost of Eq. (\ref{eq:cost}):
\begin{equation}
\label{eq:min_cost}
    S_f = \min_{P \in \mathcal{P}} \mathcal{C}(f,P)
\end{equation}

Intuitively, when $\alpha$ increases, a greater structural constraint is applied to find the closest prototype to $f$. If no similar prototype in terms of feature representation can be found in the spatial neighbourhood of the embedding, the assignment will either favour a far prototype with higher feature similarity, or a nearby prototype with lower feature similarity. In both cases, one of the two components of the cost (\ref{eq:cost}) will increase, leading to a higher anomaly score $S_f$ compared to normal embeddings, which are able to match with semantically and spatially coherent prototypes.

This process yields an anomaly map of size $H_l \times W_l$ for each scale $l$, computed separately using global ($\alpha=0$) and local ($\alpha > 0$) prototypes. The two resulting maps are then averaged and upsampled to the original image resolution to produce the anomaly map $A_l$. The final prediction $\mathcal{A}$ for anomaly localization is obtained by averaging the predictions across all scales $\mathcal{A} = \sum_{l} A_l$. The image-level score is defined as the maximum value of the final anomaly map.

\section{Experiments and Discussions}
\subsection{Experimental Setup}

\subsubsection{Datasets.}
 We use two public datasets for anomaly detection and localization on industrial images: MVTec AD \cite{bergmann2019_mvtecad} and MVTec AD LOCO \cite{Bergmann2022_mvtecadloco}. The first comprises 15 categories, which can either be objects or textures, each containing around 100 to 300 anomaly-free images per category for the training and between 44 and 175 images in the test set with various types of anomalies. The second was introduced more recently to propose more diverse defect types. The images are divided into five categories, each with around 400 samples for the training and 300 for the evaluation. The test samples contain anomalies that can be either structural defects or logical inconsistencies within the images.

\subsubsection{Metrics.}
On MVTec AD, performance in anomaly segmentation is measured with the pixel-level AU-ROC and the image-level detection performance is measured with the AU-ROC based on the predicted image-level anomaly score. On MVTec AD LOCO, following \cite{Bergmann2022_mvtecadloco}, we use the AU-sPRO metric up to a false positive rate of 5\% to assess the localization performance, and the AU-ROC for image-level classification performance.

\subsubsection{Implementation details and hyperparameters.}

All images are resized to a common shape of $224 \times 224$. We use ResNet50 as backbone and extract intermediate features at the stages $l = 2,3$, resulting in feature grids of respective sizes of $28 \times 28$ and $14 \times 14$. We set the number of prototypes per element of the spatial grid $n$ to 16, the exponential moving average learning rate $\eta$ to 0.95 and the value of $\alpha$ to $0.3$ for local prototypes. The weights of the prototypes are initially randomly initialized with a Gaussian distribution with a mean of 0 and a standard deviation of 1. The regularization parameter to solve the optimal transport problem (\ref{eq:ot_entropy}) is set to 0.01, and we limit the number of iterations for the Sinkhorn-Knopp algorithm to 100. The model is trained for 50 epochs with a batch size of 64 and no data augmentation is applied.


\subsection{Results}

\begin{figure}[htb!]
    \centering
    \includegraphics[width=\linewidth]{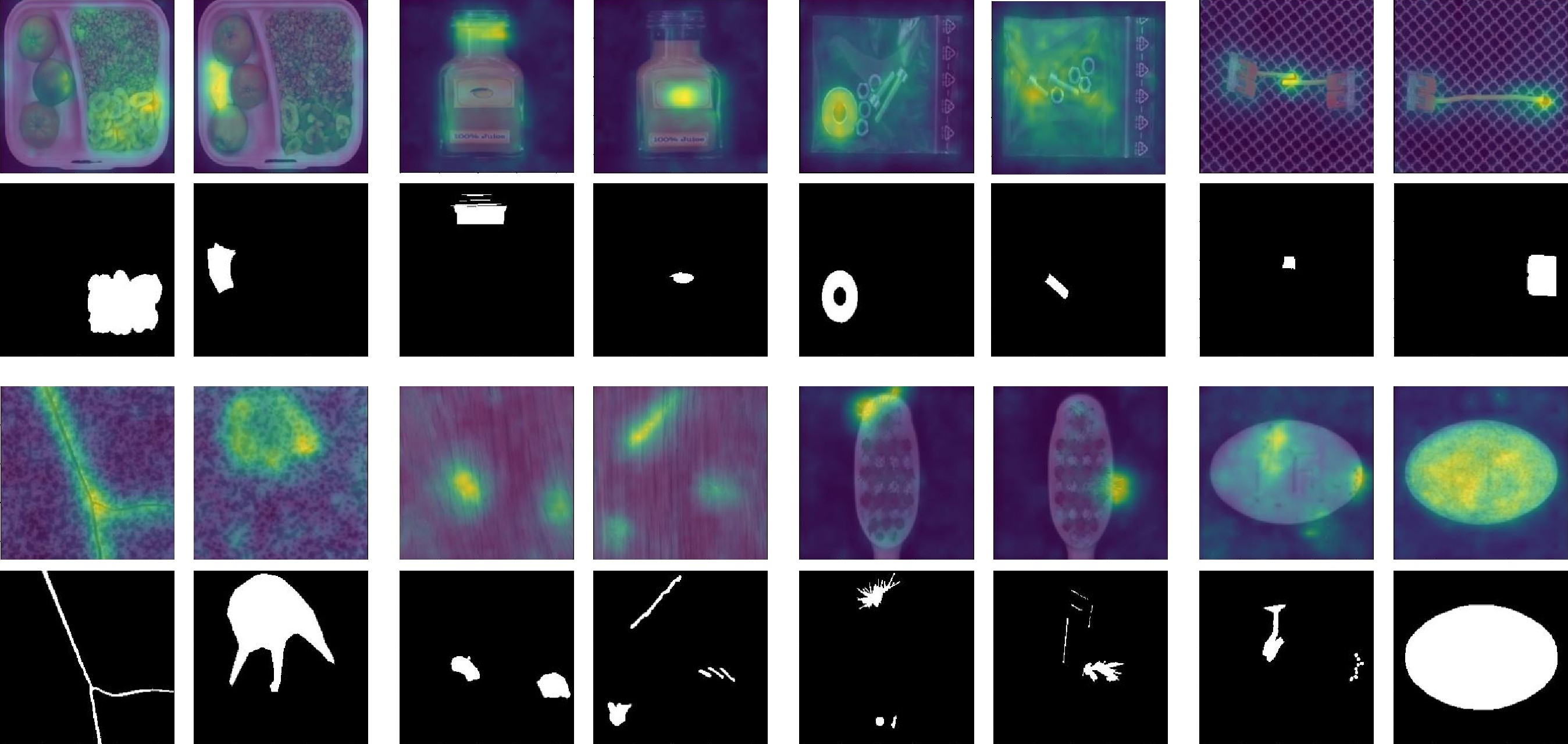}
    \caption{Examples of predictions of anomaly localization, including failure cases, on various objects of MVTec AD LOCO (top) and MVTec AD (bottom).}
    \label{fig:anomaps}
\end{figure}

We compare our approach with several baselines from the literature, namely CutPaste \cite{li2021cutpaste}, STPM \cite{wang2021studentteacherpyramid}, PaDiM \cite{brendel2018_padim}, FastFlow \cite{yu2021fastflow}, DRAEM \cite{zavrtanik2021draem}, PatchCore \cite{roth2022_patchore}, SPADE \cite{cohen2020_spade} and RD4D \cite{Deng2022_reverseKD}. For all these methods, we report their performances from the study \cite{IMIAD}, which provides an extensive and uniform benchmark on industrial image anomaly detection and localization.
Table \ref{tab:main_results} reports the results of our method compared to baselines on the datasets MVTec AD LOCO and MVTec AD, respectively. Figure \ref{fig:anomaps} showcases some visual examples of anomaly localizations for both datasets.

\begin{table}[htb!]
\caption{Results of our method compared to several UAD baselines on MVTec AD LOCO and MVTec AD. The metrics reported are the averages over all categories of each dataset. The results of the baselines are taken from \cite{IMIAD}. Full results by category are provided in Appendix \ref{app:full_results}.}
\label{tab:main_results}
\scriptsize
\setlength{\tabcolsep}{4pt}
\def\arraystretch{1.3}
\centering
\begin{tabular}{@{}lcccc@{}}
\cmidrule(l){2-5}
         & \multicolumn{2}{c}{MVTec AD LOCO} & \multicolumn{2}{c}{MVTec AD}  \\ \cmidrule(l){2-3} \cmidrule(l){4-5} 
         & Classification   & Localization   & Classification & Localization \\
         & (AU-ROC)         & (AU-sPRO)      & (AU-ROC)       & (AU-ROC)    \\ \midrule
CutPaste & 82.3 & - & 91.8 & - \\
STPM & 68.0 & 42.8 & 92.4 & 95.4 \\
PaDiM & 67.1 & 42.6 & 90.8 & 96.6 \\
SPADE & 70.1 & 52.0 & 85.4 & 95.5 \\
FastFlow & 72.0 & 35.7 & 90.5 & 95.5 \\
DRAEM & 73.6 & 42.6 & 98.1 & 97.5 \\
PatchCore & 75.5 & 34.3 & 99.2 & 99.4 \\
RD4AD & 78.7 & 63.7 & 98.6 & 97.8 \\ \midrule
PRADOT (ours) & 78.9 & 52.6 & 97.3 & 97.6 \\
\end{tabular}
\end{table}

On MVTec AD LOCO, PRADOT achieves an average image-level anomaly detection AU-ROC of 78.9\% and an AU-sPRO of 52.6\% for pixel-level segmentation. Among all compared methods, our method ranks second both in terms of classification and localization. On MVTec AD, our method reaches a classification AU-ROC score of 97.3\% and a localization score of 97.6\%.
Notably, on MVTec AD LOCO, PRADOT outperforms the SOTA memory-based method PatchCore, but the latter is significantly ahead on MVTec AD. This behavior is mainly due to the differences in types of anomalies between the two datasets. Indeed, MVTec AD mainly contains structural defects such as dents and scratches, while MVTec AD LOCO also contains anomalies in the form of violations of the logical rules, such as permissible objects occurring in invalid locations. With PRADOT, we introduce local prototypes that can identify such logical anomalies, but the way we learn the prototypes seems less effective than PatchCore for structural anomalies. On MVTec AD, most objects or textures are either symmetric or uniform. Consequently, learning local prototypes has little to no advantage compared to learning global embedding representatives. This phenomenon, as well as the role of global and local prototypes in detecting structural and logical anomalies, will be further discussed in Section \ref{sec:ablation}. 

Qualitatively, Figure \ref{fig:anomaps} shows that our method can spot both large and subtle defects. On MVTec AD LOCO, PRADOT can successfully detect structural and logical anomalies, although the segmentation performances are notably better for structural anomalies (see Figure \ref{fig:graph_batchsize_and_alpha}.b)).

\subsection{Ablation study}
\label{sec:ablation}

\subsubsection{Local, global and multi-scale aggregation.}

We investigate in Table \ref{tab:global_local_multiscale} the importance of aggregating anomaly detection maps obtained from global ($\alpha=0$) or local ($\alpha>0$) embedding-to-prototype assignment and the benefits of extracting features at various stages of the image encoder. On scales $l=2$ and $l=3$, averaging global and local predictions (\textit{Avg.}) improves the performance of the model on the two datasets and both in terms of image-level classification and pixel-level detection. An even greater performance gain is achieved with multi-scale aggregation. We can also notice that the scale $l=2$ is slightly better at pixel-level anomaly localization than the scale $l=3$. 

\begin{table}[tb!]
\caption{Impact of local and global averaging (\textit{Avg.}) and multi-scale aggregation on the performance on the datasets MVTec AD LOCO and MVTec AD. \correction{Performances are given as Image AU-ROC/Pixel AU-sPRO for MVTec AD LOCO and Image AU-ROC/Pixel AU-ROC for MVTec AD.}}
\label{tab:global_local_multiscale}
\scriptsize
\setlength{\tabcolsep}{2pt}
\def\arraystretch{1.3}
\centering
\begin{tabular}{@{ }lccccccc@{ }}
\cmidrule(l){2-8}
& \multicolumn{3}{c}{$l=2$} & \multicolumn{3}{c}{$l=3$} & \multirow{2}{*}{PRADOT} \\ \cmidrule(lr){2-4} \cmidrule(lr){5-7}
& $\alpha=0$   & $\alpha=0.3$   & Avg.  & $\alpha=0$   & $\alpha=0.3$   & Avg.  &                         \\ \midrule
MVTec AD LOCO  &   75.5/47.5    &   76.1/49.1    &  76.4/49.4     &  76.3/42.4     &  75.8/42.3       &   76.4/44.0   &  78.9/52.6                       \\
MVTec AD        &  96.7/96.1     &   96.6/96.6  & 97.0/96.8 &   95.2/96.1    &   95.2/96.0      &  95.4/96.1 &    97.3/97.6   \\ \bottomrule
\end{tabular}
\end{table}

\subsubsection{Number of prototypes and batch size.}

\begin{figure}[htb!]
    \centering
    \includegraphics[width=\linewidth]{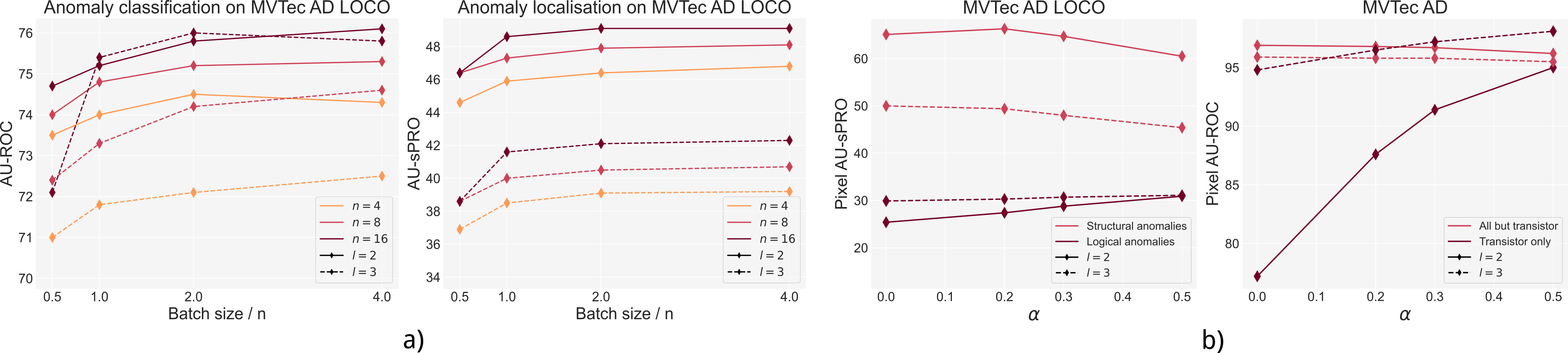}
    \caption{a) Impact of the ratio between batch size and number of prototypes per grid cell $n$ on MVTec AD LOCO in terms of image-level classification and pixel-level localization, for $\alpha=0.3$. b) Impact of the parameter $\alpha$ on the localization performance on structural and logical anomalies for MVTec AD LOCO and on all categories but \textit{transistor} and on \textit{transistor} only for MVTec AD.}
    \label{fig:graph_batchsize_and_alpha}
\end{figure}

An important hyperparameter of the model is the number of prototypes per grid cell $n$. Moreover, when learning prototypes, previous works \cite{caron2020swav} have shown that using a large batch size is essential. Therefore, we study how the parameter $n$ influences the performance of our approach and vary the ratio between the batch size and $n$ to see if a low ratio deteriorates the performance. Figure \ref{fig:graph_batchsize_and_alpha}.a) illustrates these two experiments on MVTec AD LOCO. 
As expected, performance is noticeably reduced when the ratio between the batch size and the parameter $n$ is low, in particular when the batch size is equal to or lower than $n$. However, our model does not require a very large batch size, as the improvement between a value of 2 and 4 for this ratio is small.
Regarding the number of prototypes per grid cell $n$, it plays an important role in the classification and localization performances, as we observe a substantial difference in metrics between $n=4$, $n=8$ and $n=16$. 

\subsubsection{Impact of the structural constraint.}

\begin{figure}[htb!]
    \centering
    \includegraphics[width=\linewidth]{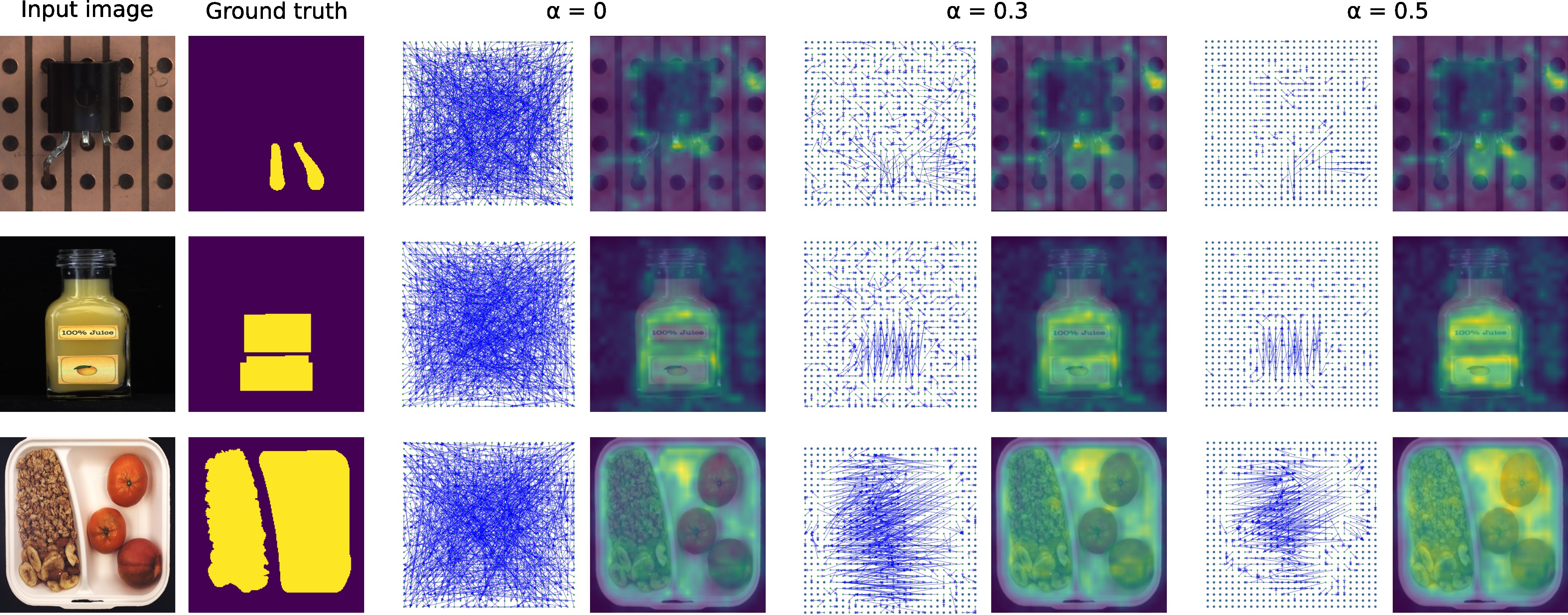}
    \caption{Examples of embedding-to-prototype assignment at scale $l=2$ for $\alpha=0, 0.3$ and $0.5$. Each blue vector starts from the position $(i,j)$ of a feature extracted from the input image and ends at the position of the prototype to which it has been assigned for the Eq. (\ref{eq:min_cost}). When $\alpha$ increases, a greater structural constraint is applied, which allows to identify some logical anomalies that global embedding-to-prototype assignments ($\alpha=0$) can not spot. \correction{Larger version of the figure available in Appendix \ref{app:assignment}.}}
    \label{fig:assignments}
\end{figure}

We investigate the impact of the structural constraint in loss (\ref{eq:cost}) and the role of global and local prototypes by evaluating how the parameter $\alpha$ influences the classification and localization performance. The results of this experiment are shown in Figure \ref{fig:graph_batchsize_and_alpha}.b).
On MVTec AD LOCO, we distinguish performance on structural and logical anomalies. There are clear correlations between the value of $\alpha$ and the localization performance of logical and structural anomalies. A higher value of $\alpha$ is beneficial for identifying logical defects while a lower value improves the localization of structural imperfections.
On MVTec AD, all objects except \textit{transistor} are symmetric and all textures are uniform. Moreover, the anomalies are mainly alterations of the shape, surface or colour of the objects or textures. Thus, imposing a structural constraint on the embedding-to-prototype assignment with a strictly positive value for $\alpha$ is of little or no use. This phenomenon is clearly observable in Figure \ref{fig:graph_batchsize_and_alpha}.b), where the pixel-level AU-ROC of all categories except \textit{transistor} is slightly decreased when $\alpha$ increases, while the same metric on \textit{transistor} only is greatly boosted by a greater structural constraint, \textit{i.e.} a high value for $\alpha$. \correction{Figure \ref{fig:assignments} showcases three examples of logical anomalies and the estimated anomaly score map of the scale $l=2$ for three values of $\alpha$. We also show with blue arrows where each feature extracted from the $H_2 \times W_2$ grid finds its closest prototypes for Eq. (\ref{eq:min_cost}). The features of anomalous regions of the images are associated with prototypes located elsewhere in the image. When $\alpha$ is low, the cost (\ref{eq:min_cost}) does not much penalize such a violation of structural organization, hence the anomaly maps for $\alpha=0$ and $\alpha=0.3$ fail to catch those logical anomalies. However, for $\alpha=0.5$, the penalty for not having found a similar prototype in the neighbourhood of the features is high enough to be able to detect the anomalous regions.} 

\subsubsection{Prototypes visualization and image reconstructions.}

\begin{figure}[htb!]
    \centering
    \includegraphics[width=0.65\linewidth]{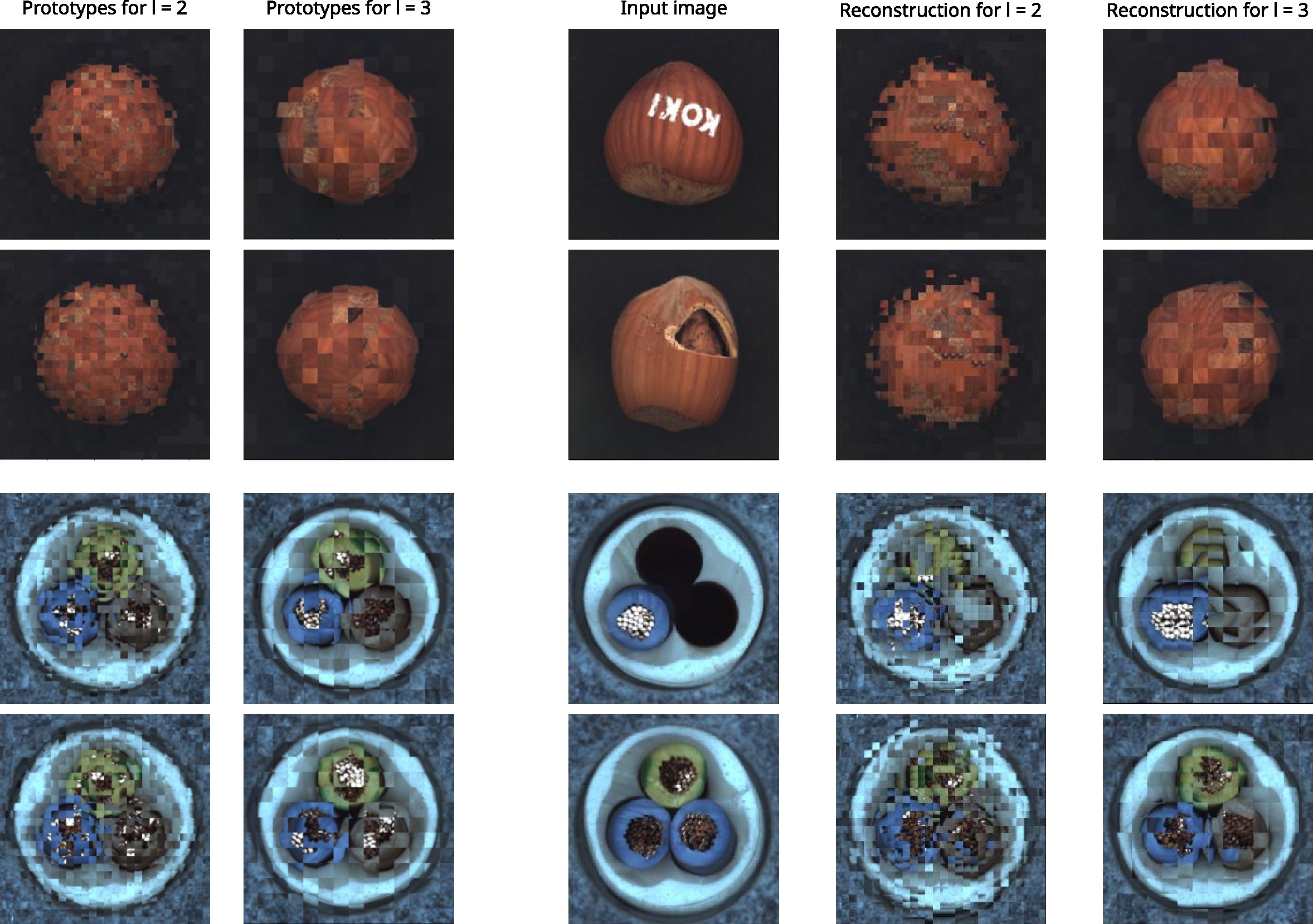}
    \caption{(left) Examples of reconstructed prototypes in the image domain by association with the patch corresponding to the feature with the highest cosine similarity. (right) For a test image, 
    we can obtain a restored version of the image where each patch comes from the representation in the pixel domain of the reconstructed prototypes.}
    \label{fig:proto_and_recons}
\end{figure}

When training is finished and the weights of the prototypes are fixed, we can try to visualize in the pixel domain what each prototype looks like. To do so, for a scale $l$ at which the features are extracted from the encoder, we divide each image of the training into a grid of $H_l \times W_l$ of non-overlapping patches of sizes $\frac{H_0}{H_l} \times \frac{W_0}{W_l}$ where $H_0$ and $W_0$ designate the height and width of an input image. For a feature map of size $H_l \times W_l$, we consider that the representation in the pixel space of a feature vector at position $(i,j)$ of the grid is the patch located at the same spot in the input image. We iterate over all the images in the training dataset and, for each prototype, we find the feature with which it maximizes the cosine similarity and associate to this prototype the patch in the pixel domain corresponding to this feature. 
The left part of Figure \ref{fig:proto_and_recons} illustrates how this procedure allows for reconstructing prototypes. We can observe that when there is more than one prototype per grid cell, \textit{i.e.} when $n>1$, each prototype can learn a different feature from the images of the training dataset.

Once this is done, when a test image is given to the model, we can look, for each feature extracted by the encoder, for the prototype which minimizes the cost $\mathcal{C}$, or the antecedent of the Eq. (\ref{eq:min_cost}), and replace in the original test image the patch corresponding to the feature by the pixel-level representation of the prototype in question. 
The right part of Figure \ref{fig:proto_and_recons} displays some examples of restored test images that can be obtained with this method. We can observe that abnormal regions are removed from the images and often replaced by semantically coherent patches. This property could potentially be leveraged to further improve the detection performance.

\subsubsection{Latent space visualization.}

\begin{figure}[htb!]
    \centering
    \includegraphics[width=0.9\linewidth]{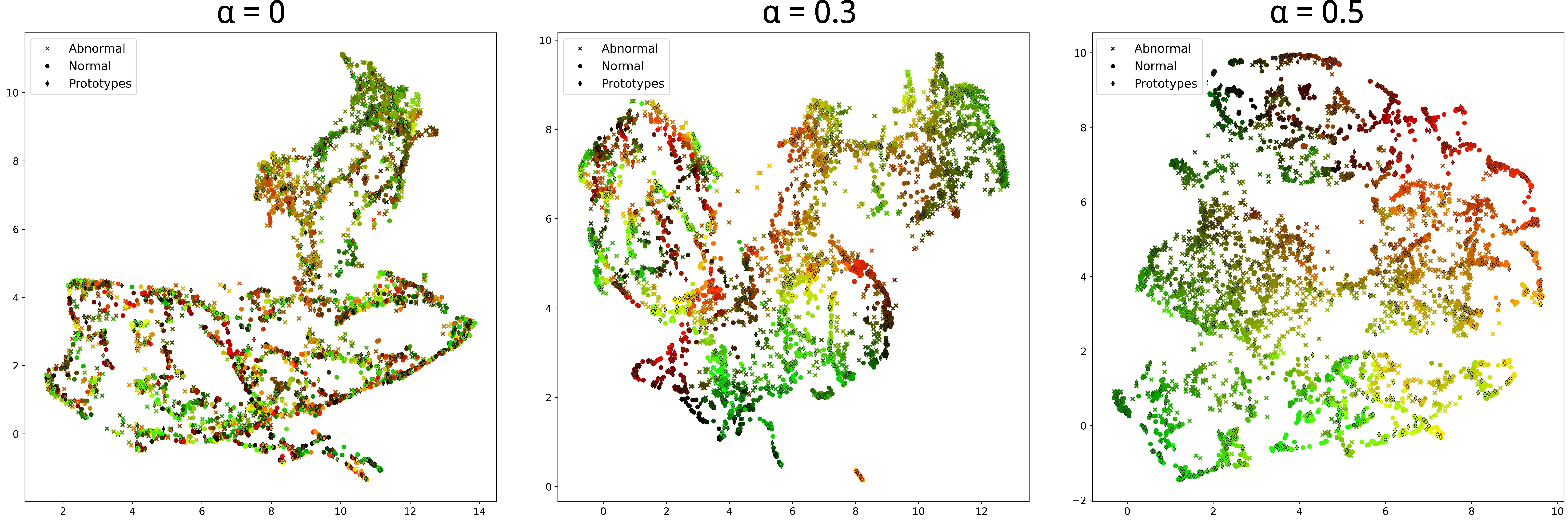}
    \caption{UMAP projections of normal embeddings, abnormal embeddings and prototypes for various values of $\alpha$. The cost (\ref{eq:cost}) is used as a pairwise distance between the points to compute their 2-dimensional projection. Colour gradient encodes the spatial position of embeddings and prototypes. \correction{Larger version of the figure available in Appendix \ref{app:umap}.}}
    \label{fig:umap}
\end{figure}

We provide with Figure \ref{fig:umap} a last visualization of the influence of the parameter $\alpha$. We project the embeddings of test images and prototypes for the object \textit{transistor} into a 2-dimensional space using UMAP \cite{2018arXivUMAP} and the cost (\ref{eq:cost}) as the pairwise distance. Each point, cross or diamond corresponds to a single embedding or prototype at the scale $l=2$. An embedding of a test image is considered abnormal if its corresponding patch in the image domain has more than 30\% of abnormal pixels. We also colour each symbol with a gradient based on the spatial location on the $H_l \times W_l$ grid of the associated embedding or prototype. This means that two embeddings or prototypes with similar colours come from the same part of the image. When $\alpha$ is close to zero, image embeddings and prototypes can be neighbouring even if they come from different regions of the images. As $\alpha$ increases, the neighbourhoods become more spatially coherent and the assignment of an embedding to a prototype must comply with this spatial constraint.

\section{Conclusion}

In this work, we propose a novel prototype-based method to tackle the challenging task of unsupervised anomaly detection on industrial images. We leverage pre-trained image encoders and optimal transport to learn prototypes. We introduce a distance function that balances a feature-based cost and a spatial-based cost. By playing on the ratio between these two components, we can enforce a structural constraint during the learning of the prototypes and at inference, which allows for better detection of incoherencies in the images. Evaluated on the public benchmarks MVTec AD LOCO and MVTec AD, our method achieves 
\correction{performance on par with strong baselines.}

One of the advantages of our approach is the simplicity and comprehensibility of the prototype learning and inference procedure. This makes it easy to understand and anticipate the strengths and weaknesses of the method depending on the application, the structure of the input images and the type of anomalies to detect. PRADOT is inspired by the problem of barycentre search in fused Gromov Wasserstein \cite{titouan2019_fgw}. Compared to our problem formulation, FGW is a graph-based approach and therefore implies different geometrical constraints, such as permutation invariance, which may be beneficial for some applications. The investigation of this extension is left for future work. We believe that the spatially consistent assignment used in our approach could have applications in other problems, such as supervised image segmentation, or combined with other UAD approaches that use a memory mechanism \cite{van2017vqvae,gong2019memoryae,guo2023_templateguided}.

\begin{credits}
\subsubsection{\ackname} \correction{This work was partially funded by  ANR-18-CE17-0012 (IMAGINA) and ANR-24-CE45-4399 (SEIZURE) grants from Agence Nationale de la Recherche (ANR). This work was granted access to the HPC resources of IDRIS under the allocation 2025-AD011014900R1 made by GENCI.}

\subsubsection{\discintname}
The authors have no competing interests to declare that are relevant to the content of this article.
\end{credits}

%

\bibliographystyle{splncs04}
\bibliography{bibliography}

\begin{thebibliography}{10}
\providecommand{\url}[1]{\texttt{#1}}
\providecommand{\urlprefix}{URL }
\providecommand{\doi}[1]{https://doi.org/#1}

\bibitem{Bergmann2022_mvtecadloco}
Bergmann, P., Batzner, K., Fauser, M., Sattlegger, D., Steger, C.: Beyond dents and scratches: Logical constraints in unsupervised anomaly detection and localization. International Journal of Computer Vision  \textbf{130}(4),  947--969 (2022)

\bibitem{bergmann2019_mvtecad}
Bergmann, P., Fauser, M., Sattlegger, D., Steger, C.: Mvtec ad — a comprehensive real-world dataset for unsupervised anomaly detection. In: 2019 IEEE/CVF Conference on Computer Vision and Pattern Recognition (CVPR). pp. 9584--9592 (2019)

\bibitem{brendel2018_padim}
Brendel, W., Bethge, M.: Approximating cnns with bag-of-local-features models works surprisingly well on imagenet. International Conference on Learning Representations  (2019)

\bibitem{cao2022otkge}
Cao, Z., Xu, Q., Yang, Z., He, Y., Cao, X., Huang, Q.: Otkge: Multi-modal knowledge graph embeddings via optimal transport. Advances in neural information processing systems  \textbf{35},  39090--39102 (2022)

\bibitem{caron2020swav}
Caron, M., Misra, I., Mairal, J., Goyal, P., Bojanowski, P., Joulin, A.: Unsupervised learning of visual features by contrasting cluster assignments. Advances in neural information processing systems  \textbf{33},  9912--9924 (2020)

\bibitem{cohen2020_spade}
Cohen, N., Hoshen, Y.: Sub-image anomaly detection with deep pyramid correspondences. arXiv preprint arXiv:2005.02357  (2020)

\bibitem{courty2016optimal_domain_adapt}
Courty, N., Flamary, R., Tuia, D., Rakotomamonjy, A.: Optimal transport for domain adaptation. IEEE transactions on pattern analysis and machine intelligence  \textbf{39}(9),  1853--1865 (2016)

\bibitem{cuturi2013_entropy}
Cuturi, M.: Sinkhorn distances: Lightspeed computation of optimal transport. In: Advances in Neural Information Processing Systems. vol.~26 (2013)

\bibitem{damodaran2018deepjdot}
Damodaran, B.B., Kellenberger, B., Flamary, R., Tuia, D., Courty, N.: Deepjdot: Deep joint distribution optimal transport for unsupervised domain adaptation. In: Proceedings of the European conference on computer vision (ECCV). pp. 447--463 (2018)

\bibitem{Deng2022_reverseKD}
Deng, H., Li, X.: Anomaly detection via reverse distillation from one-class embedding. In: Proceedings of the IEEE/CVF Conference on Computer Vision and Pattern Recognition (CVPR). pp. 9737--9746 (June 2022)

\bibitem{gong2019memoryae}
Gong, D., Liu, L., Le, V., Saha, B., Mansour, M.R., Venkatesh, S., Hengel, A.v.d.: Memorizing normality to detect anomaly: Memory-augmented deep autoencoder for unsupervised anomaly detection. In: Proceedings of the IEEE/CVF international conference on computer vision. pp. 1705--1714 (2019)

\bibitem{goodfellow2020gan}
Goodfellow, I., Pouget-Abadie, J., Mirza, M., Xu, B., Warde-Farley, D., Ozair, S., Courville, A., Bengio, Y.: Generative adversarial networks. Communications of the ACM  \textbf{63}(11),  139--144 (2020)

\bibitem{guo2023_templateguided}
Guo, H., Ren, L., Fu, J., Wang, Y., Zhang, Z., Lan, C., Wang, H., Hou, X.: Template-guided hierarchical feature restoration for anomaly detection. In: Proceedings of the IEEE/CVF International Conference on Computer Vision. pp. 6447--6458 (2023)

\bibitem{he2016resnet}
He, K., Zhang, X., Ren, S., Sun, J.: Deep residual learning for image recognition. In: Proceedings of the IEEE conference on computer vision and pattern recognition. pp. 770--778 (2016)

\bibitem{ho2020ddpm}
Ho, J., Jain, A., Abbeel, P.: Denoising diffusion probabilistic models. Advances in neural information processing systems  \textbf{33},  6840--6851 (2020)

\bibitem{izquierdo2024visualplacerecognition}
Izquierdo, S., Civera, J.: Optimal transport aggregation for visual place recognition. In: Proceedings of the ieee/cvf conference on computer vision and pattern recognition. pp. 17658--17668 (2024)

\bibitem{kim2019variationalprotoenc}
Kim, J., Oh, T.H., Lee, S., Pan, F., Kweon, I.S.: Variational prototyping-encoder: One-shot learning with prototypical images. In: Proceedings of the IEEE/CVF conference on computer vision and pattern recognition. pp. 9462--9470 (2019)

\bibitem{kim2023zegot}
Kim, K., Oh, Y., Ye, J.C.: Zegot: Zero-shot segmentation through optimal transport of text prompts. arXiv preprint arXiv:2301.12171  (2023)

\bibitem{li2021cutpaste}
Li, C.L., Sohn, K., Yoon, J., Pfister, T.: Cutpaste: Self-supervised learning for anomaly detection and localization. In: Proceedings of the IEEE/CVF conference on computer vision and pattern recognition. pp. 9664--9674 (2021)

\bibitem{lu2024mixtureot_ood}
Lu, H., Gong, D., Wang, S., Xue, J., Yao, L., Moore, K.: Learning with mixture of prototypes for out-of-distribution detection. arXiv preprint arXiv:2402.02653  (2024)

\bibitem{lu2023characterizing_ood_ot}
Lu, Y., Qin, Y., Zhai, R., Shen, A., Chen, K., Wang, Z., Kolouri, S., Stepputtis, S., Campbell, J., Sycara, K.: Characterizing out-of-distribution error via optimal transport. Advances in Neural Information Processing Systems  \textbf{36},  17602--17622 (2023)

\bibitem{2018arXivUMAP}
{McInnes}, L., {Healy}, J., {Melville}, J.: {UMAP: Uniform Manifold Approximation and Projection for Dimension Reduction}. ArXiv e-prints  (2018)

\bibitem{park2020_mnad}
Park, H., Noh, J., Ham, B.: Learning memory-guided normality for anomaly detection. In: Proceedings of the IEEE/CVF Conference on Computer Vision and Pattern Recognition. pp. 14372--14381 (2020)

\bibitem{peyre2016gromov}
Peyr{\'e}, G., Cuturi, M., Solomon, J.: Gromov-wasserstein averaging of kernel and distance matrices. In: International conference on machine learning. pp. 2664--2672. PMLR (2016)

\bibitem{roth2022_patchore}
Roth, K., Pemula, L., Zepeda, J., Sch{\"o}lkopf, B., Brox, T., Gehler, P.: Towards total recall in industrial anomaly detection. In: Proceedings of the IEEE/CVF conference on computer vision and pattern recognition. pp. 14318--14328 (2022)

\bibitem{ruff18a_deepsvdd}
Ruff, L., Vandermeulen, R.A., G{\"o}rnitz, N., Deecke, L., Siddiqui, S.A., Binder, A., M{\"u}ller, E., Kloft, M.: Deep one-class classification. In: Proceedings of the 35th International Conference on Machine Learning. vol.~80, pp. 4393--4402 (2018)

\bibitem{schlegl2019fanogan}
Schlegl, T., Seeb{\"o}ck, P., Waldstein, S.M., Langs, G., Schmidt-Erfurth, U.: f-anogan: Fast unsupervised anomaly detection with generative adversarial networks. Medical image analysis  \textbf{54},  30--44 (2019)

\bibitem{schluter2022naturalsynthetic}
Schl{\"u}ter, H.M., Tan, J., Hou, B., Kainz, B.: Natural synthetic anomalies for self-supervised anomaly detection and localization. In: European Conference on Computer Vision. pp. 474--489. Springer (2022)

\bibitem{shan_fewshotadot}
Shan, D., Zhang, Y., Coleman, S., Kerr, D., Liu, S., Hu, Z.: Unseen-material few-shot defect segmentation with optimal bilateral feature transport network. IEEE Transactions on Industrial Informatics  \textbf{19}(7),  8072--8082 (2023). \doi{10.1109/TII.2022.3216900}

\bibitem{snell2017prototypicalnetworks_fewshot}
Snell, J., Swersky, K., Zemel, R.: Prototypical networks for few-shot learning. Advances in neural information processing systems  \textbf{30} (2017)

\bibitem{tax2004svdd}
Tax, D.M., Duin, R.P.: Support vector data description. Machine learning  \textbf{54},  45--66 (2004)

\bibitem{thual2022aligning_ufgw}
Thual, A., Tran, Q.H., Zemskova, T., Courty, N., Flamary, R., Dehaene, S., Thirion, B.: Aligning individual brains with fused unbalanced gromov wasserstein. Advances in neural information processing systems  \textbf{35},  21792--21804 (2022)

\bibitem{tian2025fastref_fewshot}
Tian, L., Li, Y., Dai, Y., Chen, W., Liu, X., Chen, B.: Fastref:fast prototype refinement for few-shot industrial anomaly detection (2025), \url{https://arxiv.org/abs/2506.21398}

\bibitem{van2017vqvae}
Van Den~Oord, A., Vinyals, O., et~al.: Neural discrete representation learning. Advances in neural information processing systems  \textbf{30} (2017)

\bibitem{titouan2019_fgw}
Vayer, T., Courty, N., Tavenard, R., Laetitia, C., Flamary, R.: Optimal transport for structured data with application on graphs. In: Proceedings of the 36th International Conference on Machine Learning. vol.~97, pp. 6275--6284 (2019)

\bibitem{vincent2022template_graph_ot}
Vincent-Cuaz, C., Flamary, R., Corneli, M., Vayer, T., Courty, N.: Template based graph neural network with optimal transport distances. Advances in Neural Information Processing Systems  \textbf{35},  11800--11814 (2022)

\bibitem{wang2021studentteacherpyramid}
Wang, G., Han, S., Ding, E., Huang, D.: Student-teacher feature pyramid matching for anomaly detection. arXiv preprint arXiv:2103.04257  (2021)

\bibitem{IMIAD}
Xie, G., Wang, J., Liu, J., Lyu, J., Liu, Y., Wang, C., Zheng, F., Jin, Y.: Im-iad: Industrial image anomaly detection benchmark in manufacturing. IEEE Transactions on Cybernetics  \textbf{54}(5),  2720--2733 (2024)

\bibitem{xu2024ot_unsupervisedactionseg}
Xu, M., Gould, S.: Temporally consistent unbalanced optimal transport for unsupervised action segmentation. In: Proceedings of the IEEE/CVF Conference on Computer Vision and Pattern Recognition. pp. 14618--14627 (2024)

\bibitem{xu2023multimodal_ot_mil}
Xu, Y., Chen, H.: Multimodal optimal transport-based co-attention transformer with global structure consistency for survival prediction. In: Proceedings of the IEEE/CVF international conference on computer vision. pp. 21241--21251 (2023)

\bibitem{yan2021maskgan}
Yan, X., Zhang, H., Xu, X., Hu, X., Heng, P.A.: Learning semantic context from normal samples for unsupervised anomaly detection. In: Proceedings of the AAAI conference on artificial intelligence. vol.~35, pp. 3110--3118 (2021)

\bibitem{yao2024glad}
Yao, H., Liu, M., Yin, Z., Yan, Z., Hong, X., Zuo, W.: Glad: towards better reconstruction with global and local adaptive diffusion models for unsupervised anomaly detection. In: European Conference on Computer Vision. pp. 1--17. Springer (2024)

\bibitem{yi2020patchsvdd}
Yi, J., Yoon, S.: Patch svdd: Patch-level svdd for anomaly detection and segmentation. In: Proceedings of the Asian conference on computer vision (2020)

\bibitem{yu2021fastflow}
Yu, J., Zheng, Y., Wang, X., Li, W., Wu, Y., Zhao, R., Wu, L.: Fastflow: Unsupervised anomaly detection and localization via 2d normalizing flows. arXiv preprint arXiv:2111.07677  (2021)

\bibitem{zavrtanik2021draem}
Zavrtanik, V., Kristan, M., Sko{\v{c}}aj, D.: Draem-a discriminatively trained reconstruction embedding for surface anomaly detection. In: Proceedings of the IEEE/CVF international conference on computer vision. pp. 8330--8339 (2021)

\bibitem{zavrtanik2021reconstructionlimit}
Zavrtanik, V., Kristan, M., Sko{\v{c}}aj, D.: Reconstruction by inpainting for visual anomaly detection. Pattern Recognition  \textbf{112},  107706 (2021)

\bibitem{zhang2024contextualknowledgedistillation}
Zhang, J., Suganuma, M., Okatani, T.: Contextual affinity distillation for image anomaly detection. In: Proceedings of the IEEE/CVF Winter Conference on Applications of Computer Vision. pp. 149--158 (2024)

\bibitem{zhou2022rethinking_proto}
Zhou, T., Wang, W., Konukoglu, E., Van~Gool, L.: Rethinking semantic segmentation: A prototype view. In: Proceedings of the IEEE/CVF conference on computer vision and pattern recognition. pp. 2582--2593 (2022)

\bibitem{zhou2024_msflow}
Zhou, Y., Xu, X., Song, J., Shen, F., Shen, H.T.: Msflow: Multiscale flow-based framework for unsupervised anomaly detection. IEEE Transactions on Neural Networks and Learning Systems  (2024)

\end{thebibliography}

\clearpage

\appendix

\section{Additional results}
\label{app:full_results}

\begin{table}[htb!]
\caption{Results of our method compared to several UAD baselines on all five categories of MVTec AD LOCO. Performance metrics are displayed as: Image-level AU-ROC / Pixel-level AU-sPRO (5\%). Results of the baselines are taken from \cite{IMIAD}.}
\label{tab:mvtecad_loco}
\scriptsize
\setlength{\tabcolsep}{2pt}
\def\arraystretch{1.5}
\centering
\begin{tabular}{@{ }lcccccc@{}}
\toprule
Method             & Breakfast box & Juice bottle & Pushpins  & Screw bag & Connectors & Mean      \\ \midrule
CutPaste & 86.4/- & 92.5/- & 72.6/- & 78.0/- & 81.9/- & 82.3/- \\
STPM & 50.7/14.2 & 85.2/65.4 & 68.8/43.5 & 67.6/34.8 & 67.9/56.3 & 68.0/42.8 \\
PaDiM & 57.5/49.4 & 88.1/82.6 & 63.2/32.6 & 61.5/45.8 & 65.1/50.0 & 67.1/52.1 \\
FastFlow & 59.3/12.2 & 84.7/52.9 & 66.4/37.7 & 72.2/14.5 & 77.1/61.4 & 72.0/35.7 \\
DRAEM & 73.4/44.2 & 86.9/73.3 & 69.1/39.4 & 71.8/20.7 & 66.9/35.3 & 73.6/42.6\\
PatchCore & 77.7/38.1 & 86.1/39.4 & 70.2/22.3 & 74.8/45.1 & 68.7/26.3 & 75.5/34.3 \\
SPADE & 63.0/24.3 & 88.0/73.4 & 62.6/42.7 & 60.9/50.4 & 75.9/68.9 & 70.1/52.0 \\
RD4AD & 68.7/42.3 & 93.9/83.8 & 74.5/62.1 & 71.2/57.4 & 85.3/73.0 & 78.7/63.7 \\
PRADOT (ours) & 81.6/56.9 & 89.7/68.9 & 73.0/39.0 & 70.5/44.4 & 79.6/53.9 & 78.9/52.6 \\
\bottomrule
\end{tabular}
\end{table}

\begin{table}[htb!]
\caption{Results of our method compared to several UAD baselines on all fifteen categories of MVTec AD. The performance metrics are displayed as: Image-level AU-ROC / Pixel-level AU-ROC. The results of the baselines are taken from \cite{IMIAD}.}
\label{tab:mvtecad}
\scriptsize
\setlength{\tabcolsep}{2pt}
\def\arraystretch{1.5}
\centering
\begin{tabular}{{@{ }lccccccc@{ }}}
\toprule
Method & SPADE & DRAEM & FastFlow & PaDiM & PatchCore & STPM & PRADOT (ours) \\ \midrule
Carpet & 95.0/99.2 & 96.3/96.2 & 87.5/96.1 & 88.9/98.9 & 98.8/99.2 & 92.5/98.8 & 99.9/99.2 \\
Grid & 91.3/84.6 & 100.0/99.6 & 88.1/95.9 & 90.4/96.5 & 97.9/99.0 & 77.4/76.5 & 96.2/97.5 \\
Leather & 94.9/98.3 & 100.0/98.9 & 96.1/97.5 & 84.3/98.1 & 100.0/99.4 & 95.3/98.6 & 100.0/99.3 \\
Tile & 97.1/98.9 & 100.0/99.5 & 63.2/92.3 & 86.5/97.8 & 99.5/96.4 & 99.0/98.4 & 100.0/95.7 \\
Wood & 72.2/95.1 & 99.2/97.2 & 90.8/94.3 & 80.8/95.7 & 99.1/94.9 & 96.3/98.8 & 99.0/94.6 \\ \midrule
Bottle & 99.4/97.1 & 96.9/99.3 & 100.0/98.0 & 99.4/98.4 & 100.0/98.8 & 100.0/98.3 & 100.0/98.5 \\
Cable & 86.2/91.6 & 93.4/95.4 & 91.2/93.5 & 85.0/95.4 & 99.7/98.8 & 74.8/88.4 & 99.1/98.4 \\
Capsule & 97.3/85.8 & 96.1/94.0 & 99.2/93.0 & 94.9/90.6 & 97.9/99.2 & 94.4/96.9 & 93.4/98.4 \\
Hazelnut & 88.7/92.5 & 100.0/99.5 & 98.8/92.7 & 99.1/93.0 & 100.0/99.0 & 95.6/87.8 & 100.0/98.9 \\
Metal & 83.4/98.7 & 99.4/98.7 & 90.8/97.4 & 87.6/98.5 & 99.9/98.7 & 94.0/98.0 & 100.0/96.6 \\
Pill & 78.4/97.5 & 96.8/97.6 & 98.0/98.9 & 98.1/98.8 & 96.7/98.3 & 93.0/98.7 & 95.7/97.2 \\
Screw & 47.4/97.6 & 99.2/99.7 & 97.8/98.1 & 92.2/94.5 & 98.8/99.6 & 97.6/98.9 & 91.0/97.9 \\
Toothbrush & 90.5/99.2 & 99.7/98.1 & 100.0/99.4 & 100.0/99.2 & 100.0/98.8 & 100.0/99.4 & 88.3/98.6 \\
Transistor & 85.9/97.1 & 94.2/90.0 & 88.1/92.7 & 94.1/96.6 & 99.9/96.1 & 97.2/95.5 & 99.3/96.2 \\
Zipper & 73.3/99.4 & 99.7/98.6 & 67.5/93.4 & 80.9/97.8 & 99.5/99.0 & 78.5/97.6 & 97.2/96.8 \\ \midrule
Average & 85.4/95.5 & 98.1/97.5 & 90.5/95.5 & 90.8/96.6 & 99.2/99.4 & 92.4/95.4 & 97.3/97.6 \\
\bottomrule
\end{tabular}
 \end{table}

\section{Pseudocode for prototype learning algorithm}
\label{app:pseudo_code}
\RestyleAlgo{ruled}
\begin{algorithm}
\caption{Prototype learning with PRADOT}\label{algo:proto_learning}
\DontPrintSemicolon
\SetKwInput{kwInput}{Input}
\SetKwInput{kwInit}{Initialization}
\kwInput{ \\
$\mathcal{D}^{\text{train}} = \{ x_i \}_{i=1}^{N_{\text{train}}}$, pre-trained feature encoder $\phi_{\theta}$, stages $\{l\}_{l \in L} $ at which features are extracted from $\phi_{\theta}$, number of prototypes per grid cell $n$, cost parameter $\alpha$, EMA rate $\eta$, entropic regularization $\epsilon$, batch size $B$.
}
\kwInit{ \\
\ForEach{intermediate stage $l \in L$}{
Prototypes $\mathcal{P}_l = \{(p_{i,l}, \rho_{i,l})\}_{i=1}^{n \times H \times W}$.
}
}

\BlankLine
\While{not converged}{
    Sample $\{x_k\}_{k=1}^B$ from $\mathcal{D}^{train}$ \\
    Extract multi-scale features: $ \{z_k\}_{l \in L} \gets \phi_\theta(x_k)$ \\
    
    \ForEach{intermediate stage $l \in L$}{
        Compute cost matrix $M$ (Eq.\ref{eq:cost_M}) between embeddings and prototypes \\
        Solve entropic OT problem (Eq.\ref{eq:ot_entropy}) with Sinkhorn algorithm to obtain $T^*$ \\
        \For{prototype $p_i$}{
            Update $p_{i,l} \gets \eta p_{i,l} + (1-\eta) (n \times H_l \times W_l) \sum_{k=1}^{B\times H_l \times W_l} T^*(k,i) z_k$
        }
    }
}
\end{algorithm}

\newpage

\section{Embedding-to-prototype assignment}
\label{app:assignment}
\begin{figure}[htbp!]
    \centering
    \includegraphics[height=0.87\textheight]{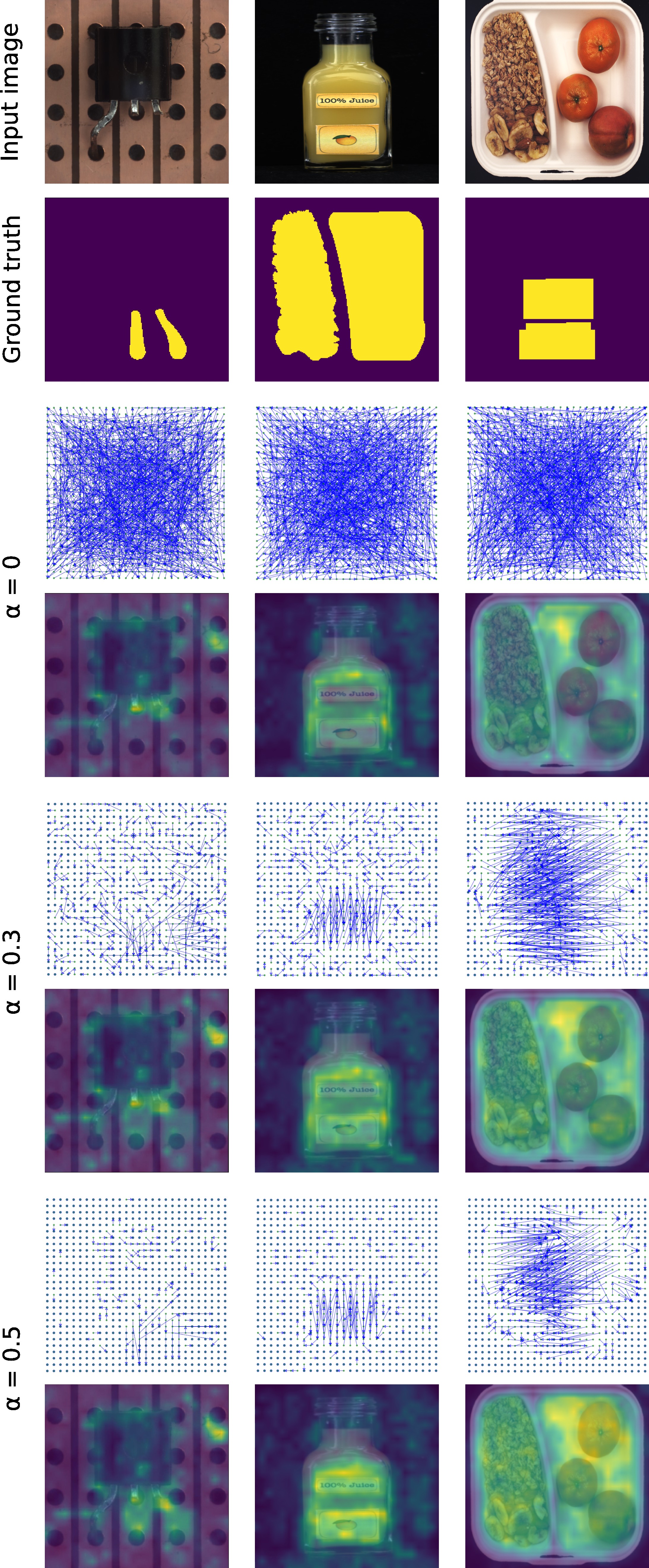}
    \caption{Bigger version of Figure \ref{fig:assignments}. Best viewed zoomed in.}
    \label{fig:appendix_assignment}
\end{figure}

\newpage

\section{UMAP projection of embeddings}
\label{app:umap}
\begin{figure}[htbp!]
    \centering
    \includegraphics[height=0.87\textheight]{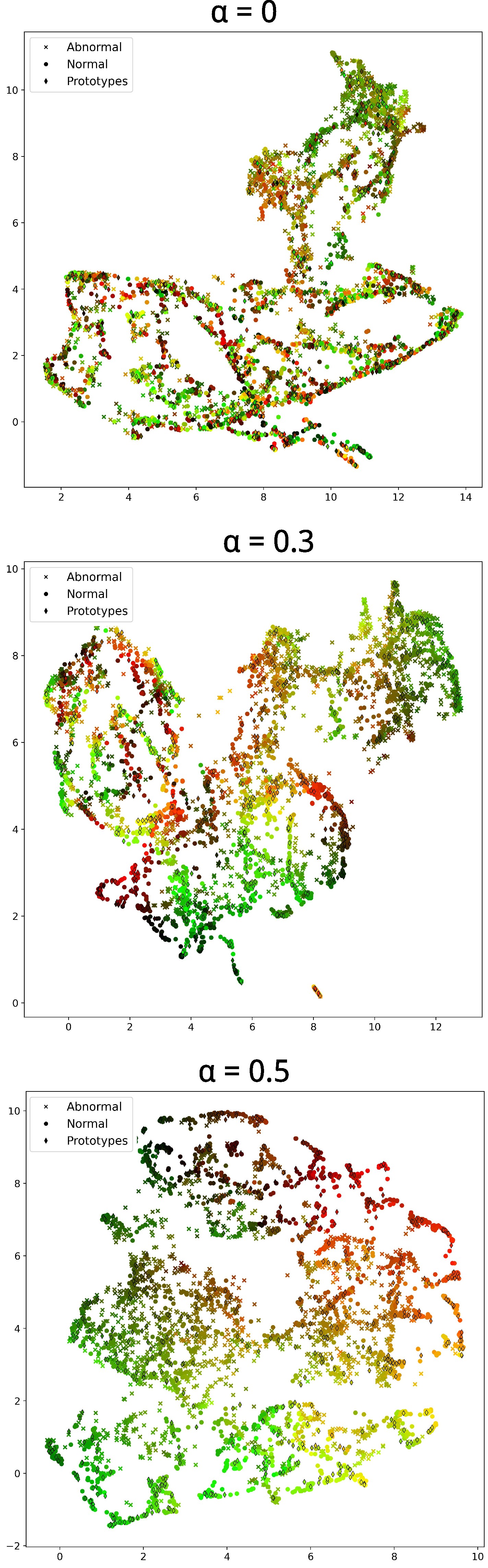}
    \caption{Bigger version of Figure \ref{fig:umap}. Best viewed zoomed in.}
    \label{fig:appendix_assignment}
\end{figure}

\end{document}